\documentclass{article}
\usepackage[cmex10]{amsmath}
\usepackage{amssymb}
\usepackage{graphicx} 
\usepackage{url}
\usepackage{changepage}
\usepackage{algorithm}
\usepackage{algorithmic}

\newtheorem{definition}{Definition}{\bfseries}{\itshape}
{\bfseries}{\itshape}

\begin{document}

\title{New Directions in Anonymization: Permutation Paradigm, Verifiability
by Subjects and Intruders, Transparency to Users}

\author{Josep Domingo-Ferrer\thanks{
Josep Domingo-Ferrer is
with the UNESCO Chair in Data Privacy,
Department of Computer Engineering and Mathematics,
Universitat Rovira i Virgili,
Av. Pa\"{\i}sos Catalans 26, E-43007 Tarragona, Catalonia,
e-mail josep.domingo@urv.cat.}
and Krishnamurty Muralidhar\thanks{
Krishnamurty Muralidhar is with the 
Dept. of Marketing and Supply Chain Management,
University of Oklahoma,
307 West Brooks, Adams Hall Room 10, Norman OK 73019-4007, USA,
e-mail krishm@ou.edu.}}

{

\maketitle

\begin{abstract}
There are currently two approaches to anonymization: ``utility first'' (use an anonymization method with suitable utility features, then empirically evaluate the disclosure risk and, if necessary, reduce the risk by possibly sacrificing some utility) or ``privacy first'' (enforce a target privacy level via a privacy model, {\em e.g.}, $k$-anonymity or 
$\varepsilon$-differential privacy, without regard to utility). To get 
formal privacy guarantees, the second approach must be followed, but 
then data releases with no utility guarantees are obtained. 
Also, in general it is unclear how verifiable is anonymization 
by the data subject (how safely released
is the record she has contributed?), what type of intruder is 
being considered (what does he know and want?) 
and how transparent is anonymization 
towards the data user (what is the user told about methods 
and parameters used?).

We show that, using a generally applicable 
reverse mapping transformation,
any anonymization for microdata can be viewed as a permutation 
plus (perhaps) a small amount of noise; permutation is thus
shown to be the essential principle underlying any anonymization
of microdata, which allows
giving simple utility and privacy metrics. 
From this permutation paradigm,
a new privacy model naturally follows, which we call $(d,{\bf v})$-permuted privacy. 
The privacy ensured by this method can be verified by each subject
contributing an original record (subject-verifiability)
and also at the data set level by the data protector.
We then proceed to define a maximum-knowledge intruder model,
which we argue should be the one considered in anonymization.
Finally, we make the case for anonymization transparent to the data user, that is,
compliant with Kerckhoff's assumption (only the randomness used, if any,
must stay secret).

\end{abstract}

\section{Introduction}

In the information society, public administrations and enterprises are increasingly collecting, exchanging and releasing large amounts of sensitive and heterogeneous information on individual subjects. Typically,
a small fraction of these data is made available to the general public 
(open data) for the purposes of improving transparency, planning, business opportunities and general well-being. Other data sets are released only to scientists for research purposes, or exchanged among companies~\cite{Daries14}.

Privacy is a fundamental right included in Article 12 of the Universal Declaration of Human Rights. However, if privacy is understood as
 seclusion~\cite{Warren90}, it is hardly compatible with the information society and with current pervasive data collection. 
A more realistic notion of privacy in our time is informational self-determination. This right was mentioned for the first time in a German constitutional ruling dated 15 Dec. 1983 as ``the capacity of the individual to determine 
in principle the disclosure and use of his/her personal data'' and it
 also underlies the classical privacy definition by~\cite{Westin67}. 

Privacy legislation in most developed countries forbids releasing and/or exchanging data that are linkable to individual subjects (re-identification disclosure) or allow inferences on individual subjects (attribute disclosure). 
Hence, in order to forestall any disclosure on individual subjects, data that are intended for release and/or exchange should first undergo a process of data anonymization, sanitization, or statistical disclosure control 
({\em e.g.}, see~\cite{Hundepool12} for a reference work).

Statistical disclosure control (SDC) takes care of respondent/subject privacy by anonymizing three types of outputs: tabular data, interactive databases and microdata files. Microdata files consist of records each of which contains data about one individual subject (person, enterprise, etc.) and the 
other two types of output can be derived from microdata. 
Hence, we will focus on microdata. The usual setting in microdata SDC is for a data protector (often the same entity that owns and releases the data) to hold the original data set (with the original responses by the subjects) and modify it to reduce the disclosure risk. There are two approaches for disclosure risk control in SDC:
\begin{itemize}
\item {\em Utility first}. An anonymization method with a heuristic 
parameter choice and with suitable utility preservation
properties\footnote{It is very difficult, if not impossible, to assess 
utility preservation for all potential analyses that can be performed
on the data. Hence, by utility preservation we mean preservation of some
preselected target statistics (for example means, variances, correlations, classifications or 
even some model fitted to the original data that should be preserved by 
the anonymized data).} is 
run on the microdata set and, after that, the risk of disclosure is measured. 
For instance, the risk of re-identification can be estimated empirically by attempting record linkage between the original and the anonymized data sets (see~\cite{Torra03}), or analytically by using generic measures ({\em e.g.}, \cite{Lambert93}) or measures tailored to a specific anonymization method 
({\em e.g.}, \cite{Elamir06} for sampling). If the extant risk is deemed too high, the anonymization method must be re-run with more privacy-stringent parameters and probably with more utility sacrifice.
\item {\em Privacy first}. In this case, a privacy model is enforced with a parameter that guarantees an upper bound on the re-identification disclosure risk and perhaps also on the attribute disclosure risk. Model enforcement is achieved by using a model-specific anonymization method with parameters that derive from the model parameters. Well-known privacy models include $\varepsilon$-differential privacy~\cite{Dwork06}, $\varepsilon$-indistinguishability~\cite{Dworket06}, $k$-anonymity~\cite{Samarati98} and the extensions of the latter taking care of attribute disclosure, like $l$-diversity~\cite{Machanavajjhala07}, 
$t$-closeness~\cite{Li07}, $(n,t)$-closeness~\cite{Li10}, crowd-blending 
privacy~\cite{Gehrke12} and others. If the utility of the resulting anonymized 
data is too low, then the privacy model in use should be enforced with a less strict privacy parameter or even replaced by a different privacy model.
\end{itemize}

\subsection{Diversity of anonymization principles}

Anonymization methods for microdata rely on a diversity of principles,
and this makes it difficult to analytically compare
their utility and data protection properties~\cite{Duncan91}; 
this is why one usually resorts to empirical 
comparisons~\cite{Domingo-Ferrer01}.
A first high-level distinction is between 
data masking and synthetic data generation. 
Masking generates a modified version ${\bf Y}$ of the original data microdata set ${\bf X}$, and it can be perturbative masking (${\bf Y}$ is a perturbed version of the original microdata set ${\bf X}$) or non-perturbative masking 
(${\bf Y}$ is obtained from ${\bf X}$ by partial suppressions or reduction of detail, yet the data in ${\bf Y}$ are still true). Synthetic data are 
artificial ({\em i.e.} simulated) data ${\bf Y}$ that preserve some preselected properties of the original data ${\bf X}$. The vast majority of anonymization methods are {\em global methods}, in that a data protector with access to the full original data set applies the method and obtains the anonymized data
set. There exist, however, {\em local perturbation methods}, in which the subjects do not need to trust anyone and
can anonymize their own data ({\em e.g.}, \cite{Rastogi07,Song14}).

\subsection{Shortcomings related to subjects, intruders and users}

We argue that current anonymization practice does not 
take the informational self-determination of the subject into account.
Since in most cases the data releaser is held legally responsible for the anonymization (for example, this happens in official statistics), the releaser favors global anonymization methods, where he can make all choices (methods, parameters, privacy and utility levels, etc.). When supplying their data, the subjects must hope there will be a data protector who will adequately protect their privacy in case of release. 
Whereas this hope may be reasonable for government surveys, it may be less so for private surveys (customer satisfaction surveys, loyalty program questionnaires, social network profiles, etc.). Indeed, a lot of privately collected data sets end up in the hands of data 
brokers~\cite{FTC14}, who trade with them with little or no 
anonymization. Hence, 
there is a fundamental mismatch between the kind of subject privacy (if any) offered by data releasers/protectors and privacy understood as informational self-determination.

The intruder model is also a thorny issue in anonymization.
In the utility-first approach and in privacy models belonging to
the $k$-anonymity family, 
restrictive assumptions are made on the amount of 
background knowledge available to the intruder for re-identification.
Assuming that a generic intruder knows this but not that is often
rather arbitrary.
In the $\varepsilon$-differential privacy model, no restrictions
are placed on the intruder's knowledge; the downside is that, 
to protect against re-identification by such an intruder,
the original data set must be perturbed to an extent such
that the presence or absence of {\em any} particular original record
becomes unnoticeable in the anonymized data set.
How to deal with an unrestricted intruder incurring as little
utility damage as possible is an open issue.

Another unresolved debate is how much detail shall or can be given to the user
 on the masking methods and parameters used to anonymize a data
 release~\cite{Cox11}. Whereas the user would derive increased inferential 
utility from learning as much as possible on how anonymization was performed, 
for some methods such details might result in disclosure of original data. 
Thus, even though Kerckhoff's principle is considered a golden rule in data encryption (encryption and decryption algorithms must be public and the only secret parameter must be the key), it is still far from being achieved/accepted in data anonymization.

\subsection{Contribution and plan of this paper}

We first give in Section~\ref{reverse} a 
procedure that, for any anonymization method,
allows mapping the anonymized attribute values back to the 
original attribute values, thereby preserving the marginal 
distributions of original attributes ({\em reverse mapping}).

Based on reverse mapping, 
we show in Section~\ref{paradigm} 
that any anonymization method for microdata can
be regarded as a permutation that may be supplemented by a small
noise addition ({\em permutation paradigm}).
Permutation is thus shown to be the essential principle underlying
any anonymization of microdata, which allows giving simple 
utility and privacy metrics that can also be used to compare
methods with each other. 

From the permutation paradigm, {\em a new privacy model} naturally 
follows, which we present in Section~\ref{permuted}
under the name {\em $(d,{\bf v})$-permuted privacy}.
Like all other privacy models, this model can be verified
by the data protector for the entire original data set.
A more attractive feature
is that {\em the subject contributing each original record
can verify to what extent the privacy guarantee of the model holds for her record} 
({\em subject-verifiability}).
Note that subject-verifiability is a major step towards informational
self-determination, because it gives the subject control on 
how her data have been anonymized (a property that has also been called
intervenability~\cite{Rost09}).

Then in Section~\ref{intruder}  
we introduce {\em a maximum-knowledge
intruder model}, which makes
any assumptions about background knowledge unnecessary. 
We describe how such an intruder can optimally guess the correspondence 
between anonymized and original records and how he can
assess the accuracy of his guess.
Further, we show how to protect against
such a powerful intruder 
by using anonymization methods that provide an adequate
level of permutation.

Finally, in Section~\ref{transparency}  
we make the case for {\em anonymization
transparent to the data user}. 
Just as Kerckhoff's
assumption is the guiding principle in data encryption,
it should be adopted in anonymization: good anonymization
methods should remain safe when
everything (anonymized data, original data, anonymization method
and parameters) except
the anonymization key (randomness used) is published.

We illustrate all concepts introduced with a running example.
Finally, conclusions and future research directions
are gathered in Section~\ref{conclude}.

\section{Reverse mapping of anonymized data}
\label{reverse}

We next recall a reverse-mapping procedure, which we first gave 
in the conference paper~\cite{Muralidhar14} in another context.
Let $X=\{x_1, x_2, \cdots, x_n\}$ the values taken by attribute $X$ in the original data set. Let $Y=\{y_1, y_2, \cdots, y_n\}$ represent the anonymized version of $X$. We make no assumptions about the anonymization method used 
to generate $Y$, but we assume that the values in 
both $X$ and $Y$ can be ranked in some way\footnote{
For numerical or categorical ordinal attributes,
ranking is straightforward. Even for categorical nominal
attributes, the ranking assumption is less restrictive
than it appears, because semantic distance metrics are available
that can be used to rank them (for instance,
the marginality distance in~\cite{Domingo13,SoriaVLDB14}).}; any
ties in them are broken randomly. 
Knowledge of $X$ and $Y$ allows deriving another set 
of values $Z$ via reverse mapping, as per Algorithm~\ref{alg1}.

\begin{algorithm}[h]
\caption{\sc Reverse-mapping conversion}
\label{alg1}
\begin{algorithmic}[l]
\REQUIRE Original attribute $X=\{x_1, x_2, \cdots, x_n\}$
\REQUIRE Anonymized attribute $Y=\{y_1, y_2, \cdots, y_n\}$
\FOR{$i=1$ to $n$}
\STATE Compute $j=\mbox{Rank}(y_i)$
\STATE Set $z_i = x_{(j)}$ (where $x_{(j)}$ is the value of $X$ of rank $j$)
\ENDFOR
\RETURN $Z=\{z_1,z_2, \cdots, z_n\}$
\end{algorithmic}
\end{algorithm}

Releasing the reverse-mapped attribute $Z$ instead of $Y$
has a number of advantages:
\begin{itemize} 
\item By construction, each reverse-mapped attribute preserves 
the rank correlation between the corresponding anonymized attribute and 
the rest of attributes in the data set; hence, 
{\em reverse mapping does not damage the rank correlation 
structure of the original data set more than the 
underlying anonymization method}. 
\item In fact, $Z$ incurs less information loss than $Y$ since 
{\em $Z$ preserves
the marginal distribution of the original attribute $X$}.
\item {\em Disclosure risk can be conveniently measured by the rank order 
correlation between $X$ and $Z$} (the higher, the more risk).
\end{itemize}

In Table~\ref{tau1} we give a running example. The
original data set consists of three attributes 
$X^1$, $X^2$ and $X^3$ which have been generated
by sampling 
$N(100,10^2)$, $N(1000, 50^2)$ and $N(5000,200^2)$
distributions, respectively.
The masked data set consists of three attributes
$Y^1$, $Y^2$ and $Y^3$ obtained, respectively,
from $X^1$, $X^2$ and $X^3$ by noise addition.
The noise $E^1$ added to $X^1$ was sampled from a $N(0,5^2)$,
the noise $E^2$ added to $X^2$ from a $N(0,25^2)$ and the noise
$E^3$ added to $X^3$ from a $N(0,100^2)$.
The reverse-mapped attributes obtained using 
Algorithm~\ref{alg1}~are $Z^1$, $Z^2$ and $Z^3$, respectively.

In Table~\ref{tau1} we also give the ranks of values 
for the original and masked attributes, so that 
Algorithm~\ref{alg1} can be verified on the table. By 
way of illustration, consider the first attribute of the 
the first record.
For the first original record, $X^1=103.69$. This value
turns out to be the 10th value of $X^1$ sorted in increasing
order. After adding noise to $X^1=103.69$, we get 
the masked value $Y^1=108.18$, which is the 14th value of $Y^1$
sorted in increasing order. Then, to do the reverse mapping,
we replace $Y^1=108.18$ by the 14th value of $X^1$ (108.21) and
we get $Z^1=108.21$.  

Clearly the values of each $Z^j$ are a permutation of the 
values of the corresponding $X^j$, for $j=1,2,3$. Hence, the 
reverse-mapped attributes preserve the marginal distribution of 
the corresponding original attributes. The disclosure risk
can be measured by the rank correlations between
$X^1$ and $Z^1$ (0.722), between $X^2$ and $Z^2$ (0.844) 
and between $X^3$ and $Y^3$ (0.776).

\begin{table}
\caption{Running example. Original data set, formed
by attributes 
$X^1$, $X^2$ and $X^3$. Masked data set, formed
by attributes $Y^1$, $Y^2$ and $Y^3$ obtained
via noise addition. Reverse-mapped data set, formed
by attributes $Z^1$, $Z^2$ and $Z^3$. The notation
$(X^j)$ stands for the ranks of the values of $X^j$. Analogously
for $(Y^j)$.} 
\label{tau1}
\begin{adjustwidth}{-3cm}{}
{\footnotesize
\begin{tabular}{|c|c|c|c|c|c|c|c|c|c|c|c|c|c|c|}\hline
$X^1$&$X^2$&$X^3$ & $(X^1)$ & $(X^2)$ & $(X^3)$ & $Y^1$ & $Y^2$ & $Y^3$ & $(Y^1)$ & $(Y^2)$ & $(Y^3)$ & $Z^1$ & $Z^2$ & $Z^3$\\\hline	
103.69	& 981.80& 4928.80	&10 &8 &8	&108.18	&972.62	&4876.73	&14	&7	&5	&108.21	&980.97	&4893.50\\
93.13	&980.97	&4931.16	&2 &7 &9	&96.60	&1020.73 &5005.04	&6	&11	&13	&96.18	&988.44	&4986.25\\
100.87	&902.21	&5108.54	&9 &1 &15	&105.26	&882.92	&4900.68	&13	&1	&7	&107.62	&902.21	&4905.71\\
95.24	&953.37	&5084.18	&4 &4 &14	&88.02	&944.54	&4949.78	&2	&4	&10	&93.13	&953.37	&4941.81\\
96.18	&1086.34 &5212.25	&6 &20 &18	&91.57	&1057.83 &5267.57	&5	&18	&19	&95.50	&1052.34	&5232.96\\
93.16	&986.70	&5232.96	&3 &10 &19	&100.41	&991.34	&5230.64	&8	&9	&18	&99.72	&984.87	&5212.25\\
95.50	&952.13	&4824.95	&5 &3 &3	&100.31	&959.89	&4824.03	&7	&5	&4	&98.99	&971.09	&4835.05\\
115.53	&988.44	&5437.43	&19 &11 &20	&123.37	&1061.23	&5450.70	&20	&19	&20	&116.75	&1057.63	&5437.43\\
98.99	&941.48	&4835.05	&7 &2 &4	&103.12	&903.25	&4752.03	&10	&2	&3	&103.69	&941.48	&4824.95\\
109.96	&984.87	&4950.48	&16 &9 &11	&104.82	&912.77	&4997.61	&12	&3	&12	&105.59	&952.13	&4954.28\\
99.72	&1005.19 & 5158.64	&8 &13 &17	&87.83	&1025.01 &5166.63	&1	&12	&17	&87.62	&990.58	&5158.64\\
116.75	&1057.63 &4986.25	&20 &19 &13	&112.21	&1082.43 &4988.44	&15	&20	&11	&109.81	&1086.34	&4950.48\\
107.62	&1025.13 &4954.28	&13 &15 &12	&114.29	&988.93	&4889.75	&17	&8	&6	&110.63	&981.80	&4900.79\\
87.62	&1031.74 & 4905.71	&1 &17 &7	&90.83	&1049.58 &4902.04	&4	&15	&8	&95.24	&1025.13	&4928.80\\
109.81	&971.09	&4941.81	&15 &5 &10	&113.64	&1002.19 &5020.71	&16	&10	&14	&109.96	&986.70	&5084.18\\
110.63	&1052.34 & 4495.19	&17 &18 &1	&103.07	&1052.03 &4519.26	&9	&17	&1	&100.87	&1031.74	&4495.19\\
113.76	&972.20	& 4893.50	&18 &6 &5	&117.00	&962.84	&5087.90	&19	&6	&16	&115.53	&972.20	&5143.05\\
105.59	&1027.64 & 5143.05	&12 &16 &16	&89.43	&1049.97 &5072.79	&3	&16	&15	&93.16	&1027.64	&5108.54\\
108.21	&990.58	& 4714.76	&14 &12 &2	&115.79	&1036.10 &4662.73	&18	&13	&2	&113.76	&1005.19	&4714.76\\
104.74	&1023.96 &4900.79	&11 &14 &6	&104.00	&1037.00 &4931.99	&11	&14	&9	&104.74	&1023.96	&4931.16\\\hline
\end{tabular}}
\end{adjustwidth}
\end{table}

\section{A permutation paradigm of anonymization}
\label{paradigm}

Reverse mapping has 
the following broader conceptual implication: 
any anonymization method  
is {\em functionally equivalent} to a two-step procedure consisting
of a permutation step (mapping the original data set to the output of 
the reverse mapping procedure in Algorithm~\ref{alg1}) 
plus a noise addition step (adding the difference
between the reverse-mapped output and the anonymized data set). 

Specifically, take ${\bf X}$ to be the original data set, 
${\bf Y}$ the anonymized data set and ${\bf Z}$ the reverse-mapped data set 
(the values of each attribute in ${\bf Z}$ are a 
permutation of the corresponding attribute in ${\bf X}$). Now, conceptually,
{\em any anonymization method is functionally equivalent to doing
the following: 
i) permute the original data 
set ${\bf X}$ to obtain ${\bf Z}$; ii) add some noise to ${\bf Z}$ to 
obtain ${\bf Y}$}. 
The noise used to transform ${\bf Z}$ into ${\bf Y}$ is necessarily
small (residual) 
because it cannot change any rank:
note that, by the
construction of Algorithm~\ref{alg1}, the ranks
of corresponding values of ${\bf Z}$
and ${\bf Y}$ are the same.

Let us emphasize that the functional equivalence
described in the previous paragraph does not imply any actual change
in the anonymization method: we are simply saying that the way
the method transforms ${\bf X}$ into ${\bf Y}$ could be 
exactly mimicked 
by first permuting ${\bf X}$ and then
adding residual noise.

In this light, it seems rather obvious that protection against
re-ident\-ifica\-tion via record linkage 
comes from the permutation step in the above functional 
equivalence: 
as justified above, the  
noise addition step in the equivalence 
does not change any ranks, so any
rank change must come from the permutation step.
Thus, {\em any two anonymization methods can, however
different their actual operating principles,
be compared in terms
of how much permutation they achieve, that is, 
how much they modify ranks.}

On the other hand, to permute, one must have access to the full data set
or at least a part of it. 
Hence, local perturbation methods, which operate
locally by adding noise to each record, 
cannot guarantee a prescribed
permutation amount; if they protect against
re-identification, it is by means  of ``blind'' 
noise addition, which may be an overkill.

We illustrate the view of anonymization as permutation plus residual
noise on the running example (Table~\ref{tau2}).
First we permute each original attribute $X^j$
to obtain the corresponding $Z^j$, for $j=1,2,3$. Then we add
the noise ${E'}^j$
required to obtain $Y^j$ from the corresponding $Z^j$, for $i=1,2,3$.
It can be observed that, for $j=1,2,3$,  in general the values of $|{E'}^j|$ 
are substantially smaller than those of $|E^j|$ 
where $E^j = Y^j -X^j$ is the noise  
required to obtain $Y^j$ directly from $X^j$ for $j=1,2,3$.

\begin{table*}
\caption{Running example. View of masking as permutation 
plus (small) noise. Original attributes $X^j$ are permuted to get $Z^j$,
for $j=1,2,3$. Then noise ${E'}^j$ is added to $Z^j$ to get $Y^j$.
In general, for $j=1,2,3$,  less noise is required to obtain $Y^j$ from $Z^j$ than directly
from $X^j$ (compare the absolute values of columns ${E'}^j$ and $E^j$).}
\label{tau2}
\begin{adjustwidth}{-3.5cm}{}
{\footnotesize
\begin{tabular}{|c|c|c|c|c|c|c|c|c|c|c|c||c|c|c|}\hline
$X^1$&$X^2$&$X^3$ & $Z^1$ & $Z^2$ & $Z^3$ & ${E'}^1$ & ${E'}^2$ & ${E'}^3$ & $Y^1$ & $Y^2$ & $Y^3$ & $E^1$ & $E^2$ & $E^3$\\\hline	
103.69& 981.80& 4928.80 & 	108.21	&980.97	&4893.50  & -0.03 &-8.35 & -16.77 &108.18 &972.62 &4876.73 &4.49 &-9.18	&-52.07  \\
93.13	&980.97	&4931.16 &96.18	&988.44	&4986.25 &  0.42 & 32.29 & 18.79 &96.60  &1020.73 &5005.04 & 3.47 &39.76&	73.88\\
100.87	&902.21	&5108.54&107.62	&902.21	&4905.71 & -2.36 & -19.29 & -5.03 &105.26 &882.92 &4900.68& 4.39&	-19.29&	-207.86 \\
95.24	&953.37	&5084.18&93.13	&953.37	&4941.81 & -5.11 & -8.83 & 7.97  &88.02  &944.54 &4949.78& -7.22&	-8.83&	-134.40 \\
96.18	&1086.34 &5212.25&95.50	&1052.34	&5232.96 & -3.93 & 5.49	& 34.61  &91.57  &1057.83 &5267.57 &-4.61&	-28.51&	55.32\\
93.16	&986.70	&5232.96 & 99.72	&984.87	&5212.25 & 0.69	& 6.47	& 18.39 &100.41 &991.34 &5230.64 &7.25&	4.64&	-2.32 \\
95.50	&952.13	&4824.95&98.99	&971.09	&4835.05 & 1.32	& -11.20 & -11.02 & 100.31 & 959.89 & 4824.03  &4.81&	7.76&	-0.92\\
115.53	&988.44	&5437.43&116.75	&1057.63	&5437.43 & 6.62	& 3.60	& 13.27 &123.37 &1061.23 &5450.70 &7.84	&72.79&	13.27 \\
98.99	&941.48	&4835.05&103.69	&941.48	&4824.95 & -0.57 &-38.23 & -72.92 &103.12 &903.25 &4752.03 &4.13&	-38.23&	-83.02 \\
109.96	&984.87	&4950.48&105.59	&952.13	&4954.28 & -0.77 & -39.36 &43.33 &104.82 &912.77 &4997.61 &-5.14&	-72.10&	47.13\\
99.72	&1005.19 & 5158.64&87.62	&990.58	&5158.64 & 0.21	& 34.43 &7.99 &87.83  &1025.01 &5166.63 &-11.89	&19.82&	7.99\\
116.75	&1057.63 &4986.25&109.81	&1086.34	&4950.48 & 2.40	& -3.91	& 37.96 &112.21 &1082.43 &4988.44 &-4.54&	24.80&	2.19\\
107.62	&1025.13 &4954.28&110.63	&981.80	&4900.79 &3.66	& 7.13	& -11.04 &114.29 & 988.93 &4889.75 &6.67	&-36.20	&-64.53 \\
87.62	&1031.74 & 4905.71&95.24	&1025.13&4928.80 & -4.41 & 24.45 & -26.76 &90.83  &1049.58 &4902.04 &3.21&	17.84	&-3.67 \\
109.81	&971.09	&4941.81&109.96	&986.70	&5084.18  & 3.68 & 15.49 & -63.47 &113.64 &1002.19 &5020.71 &3.83	&31.10	&78.90 \\
110.63	&1052.34 & 4495.19&100.87	&1031.74	&4495.19 & 2.20	& 20.29	& 24.07 &103.07 &1052.03 &4519.26 &-7.56&	-0.31	&24.07\\
113.76	&972.20	& 4893.50&115.53	&972.20	&5143.05  & 1.47 & -9.36 & -55.15 &117.00 &962.84 &5087.90 &3.24&	-9.36	&194.40 \\
105.59	&1027.64 & 5143.05&93.16	&1027.64	&5108.54 & -3.73 & 22.33  & -35.75  &89.43  &1049.97 &5072.79 &-16.16&	22.33	&-70.26\\
108.21	&990.58	& 4714.76&113.76	&1005.19	&4714.76 & 2.03	& 30.91	& -52.03 &115.79 &1036.10 &4662.73 &7.58&	45.52	&-52.03  \\
104.74	&1023.96 &4900.79&104.74	&1023.96	&4931.16 & -0.74 & 13.04 & 0.83 & 104.00 &1037.00 & 4931.99 &-0.74&	13.04&	31.20\\\hline
\end{tabular}}
\end{adjustwidth}
\end{table*}

\section{A new subject-verifiable privacy model: $(d, {\bf v})$-permuted privacy}
\label{permuted}

In Section~\ref{paradigm}, we have argued that permutation
can be regarded as the essential principle of microdata anonymization.
This suggests adopting a new privacy
model focusing on permutation. Note that no privacy model
in the literature considers permutation.
Our proposal follows.

\begin{definition}[$(d,{\bf v})$-permuted privacy w.r.t. a record]
\label{def4}
Given a non-negative integer $d$ and an $m$-dimensional vector ${\bf v}$
of non-negative real numbers, an anonymized data set with $m$ attributes 
is said to satisfy $(d,{\bf v})$-permuted privacy {\em with 
respect to original record ${\bf x}$} if,
\begin{enumerate}
\item The permutation distance for ${\bf x}$ is at least $d$ 
in the following sense: given the 
anonymized attribute values $y^1_*, \cdots, y^m_*$ 
closest to the respective attribute values of ${\bf x}$,
no anonymized record $(y^1_p, \cdots, y^m_p)$ exists such 
that the ranks of $y^j_p$ and $y^j_*$ differ less than $d$ for 
all $j=1,\cdots, m$.  
\item For $1 \leq j \leq m$, if $y^j_*$ is the value of the anonymized 
$j$-th attribute $Y^j$ closest to 
the value $x^j$ of the $j$-th attribute of ${\bf x}$,
and $S^j(d)$ is the set of values of the sorted $Y^j$
whose rank differs no more than $d$ from $y^j_*$'s rank, 
then the variance of $S^j(d)$ is greater than the $j$-th 
component of ${\bf v}$.
\end{enumerate}
\end{definition}

\begin{definition}[$(d,{\bf v})$-permuted privacy for a data set]
An anonymized data set is said to satisfy $(d,{\bf v})$-permuted privacy if
it satisfies $(d, {\bf v})$-permuted privacy with respect
to all records in the original data set.
\end{definition}

The permutation distance mentioned in the first condition of 
Definition~\ref{def4} can be computed using Algorithm~\ref{alg2}.
A few comments on this algorithm follow: 
\begin{itemize}
\item For every attribute $Y^j$, the algorithm first determines the 
anonymized value $y^j_*$ closest to the original value $x^j$ of the subject.
If the anonymization is just a permutation without noise 
addition (either because the method used involves only
permutation or because the anonymized data set has been reverse-mapped
with Algorithm~\ref{alg1} using knowledge of the original data set), 
then $y^j_*=x^j$. 

\item The goal is to determine whether these most similar values 
$y^j_*$ for $1 \leq j \leq m$ have been permuted far enough from 
each other in terms of ranks. 
\end{itemize}  

The condition on variances in Definition~\ref{def4} 
ensures that there is enough diversity, 
similar to what $l$-diversity~\cite{Machanavajjhala07} adds to $k$-anonymity. 
Variances of non-numerical attributes can be computed as described in 
\cite{Domingo13}. 

\begin{algorithm}[h]
\caption{\sc Permutation distance computation for an original record}
\label{alg2}
\begin{algorithmic}[l]
\REQUIRE ${\bf x}=(x^1,\cdots,x^m)$ \COMMENT{Original record containing $m$ attribute values} 
\REQUIRE ${\bf Y}=\{(y^1_i,\cdots,y^m_i): i=1,\cdots,n\}$ 
\COMMENT{Anonymized data set containing 
$n$ records with $m$ attributes $Y^1, \cdots, Y^m$} 

\FOR{$j=1$ to $m$}
\STATE Let $y^j_*$ be the value of $Y^j$ closest to $x^j$
\STATE Sort ${\bf Y}$ by $Y^j$
\STATE Let $(y^j_*)$ be the rank (record no.) 
of $y^j_*$ in the sorted ${\bf Y}$ 
\FOR{$i=1$ to $n$}
\STATE Let $(y^j_i)$ be the rank of $y^j_i$ in the sorted ${\bf Y}$ 
\ENDFOR 
\ENDFOR
\STATE Set $d=0$
\WHILE{$\nexists (y^1_p, \cdots,y^m_p) \in {\bf Y}$ such
 that $\forall j=1,\cdots,m$ it holds that
$|(y^j_p) - (y^j_*)| \leq d$} 
\STATE $d=d+1$
\ENDWHILE
\RETURN permutation distance $d$
\end{algorithmic}
\end{algorithm}

Let us give a numerical illustration of how Algorithm~\ref{alg2} works. 
Assume that one wants to determine the permutation distance for the 
third original record of Table~\ref{tau2}, that is,
${\bf x}_3=$$(x^1_3,x^2_3,x^3_3)=$$(100.87,902.21, 5108.54)$.
The algorithm looks for the values of $Y^1$, $Y^2$
and $Y^3$ closest to $x^1_3$, $x^2_3$ and $x^3_3$, respectively.
These are  $y^1_*=100.41$, $y^2_* = 903.25$ and $y^3_*=5087.90$, shown
in boxes in Table~\ref{tau3}. The ranks of these 
values are $(y^1_*)=8$, $(y^2_*) = 2$ and $(y^3_*)=16$  (the reader 
can find them boxed in colums $(Y^1)$, $(Y^2)$ and $(Y^3)$ of Table~\ref{tau3}).
Then the algorithm looks for the record $(y^1_p,y^2_p,y^3_p)$ whose attribute ranks deviate minimally from $((y^1_*),(y^2_*),(y^3_*))=(8,2,16)$
(the rank deviations are shown in columns $d^1$, $d^2$ and $d^3$ of 
Table~\ref{tau3}). This record turns out to be the 10th anonymized
record (shown in underlined boldface in Table~\ref{tau3}) and its 
rank deviations from the anonymized attribute values 
closest to the original attribute values are 4, 1, 4, respectively. 
Hence, the permutation distance for the third original record is
$d_3=\max{\{4,1,4\}}=4$.

Regarding the condition on variances in Definition~\ref{def4},
we can compute the variances of the three anonymized attributes
restricted to the sets $S^1(4)$, $S^2(4)$ and $S^3(4)$, respectively.
For example 
$$S^1(4)=
\{96.60,91.57,100.41,100.31,103.12,$$
$$104.82,90.83,103.07,104.00\}$$
and the variance of the values of $S^1(4)$ is 24.70.
Similarly, for $S^2(4)$ and $S^3(4)$ the corresponding variances
are 896.76 and 20167.78, respectively. 
{\em Hence, the anonymized data set in Table~\ref{tau3} satisfies
$(d,{\bf v})$-permuted privacy with respect to the third original record,
with $d=4$ and ${\bf v}=(24.70, 896.76, 20167.78)$.}

\begin{table*}
\caption{Running example. Computation of the permutation 
distance for the third original
record ${\bf x}_3=(100.87, 902.21, 5108.54)$ of Table~\ref{tau2}. 
The permutation distance is $d_3=\max{\{4,1,4\}}=4$.}
\label{tau3}
\begin{adjustwidth}{-2cm}{}
{\footnotesize
\begin{tabular}{|c|c|c|c|c|c|c|c|c|c|c|c|c|c|c|}\hline
$Y^1$ & $Y^2$ & $Y^3$ & $(Y^1)$ & $(Y^2)$ & $(Y^3)$ & $d^1=|(Y^1)-8|$ & $d^2=|(Y^2)-2|$ & $d^3=|(Y^3)-16|$ & $\max_j d^j$ \\
\hline	
108.18	&972.62	&4876.73	&14	&7	&5&	6&	5&	11&	11\\
96.60	&1020.73 &5005.04	&6	&11	&13&	2&	9&	3&	9\\
105.26	&882.92	&4900.68	&13	&1	&7&	5&	1&	9&	9\\
88.02	&944.54	&4949.78	&2	&4	&10&	6&	2&	6&	6\\
91.57	&1057.83 &5267.57	&5	&18	&19&	3&	16&	3&	16\\
\framebox{100.41}	&991.34	&5230.64	&\framebox{8}	&9	&18&	0&	7&	2&	7\\
100.31	&959.89	&4824.03	&7	&5	&4&	1&	3&	12&	12\\
123.37	&1061.23&5450.70	&20	&19	&20&     12&	17&	4&	17\\
103.12	&\framebox{903.25}	&4752.03	&10	&\framebox{2}	&3	&2	&0	&13	&13\\
\underline{\bf 104.82}	&\underline{\bf 912.77}	&\underline{\bf 4997.61}	&12	&3	&12&	4&	1&	4&	\framebox{4}\\
87.83	&1025.01 &5166.63	&1	&12	&17&	7&	10&	1&	10\\
112.21	&1082.43 &4988.44	&15	&20	&11&	7&	18&	5&	18\\
114.29	&988.93	&4889.75	&17	&8	&6&	9&	6&	10&	10\\
90.83	&1049.58 &4902.04	&4	&15	&8&	4&	13&	8&	13\\
113.64	&1002.19 &5020.71	&16	&10	&14&	8&	8&	2&	8\\
103.07	&1052.03 &4519.26	&9	&17	&1&	1&	15&	15&	15\\
117.00	&962.84	&\framebox{5087.90}	&19	&6	&\framebox{16}&	11&	4&	0&	11\\
89.43	&1049.97 &5072.79	&3	&16	&15&	5&	14&	1&	14\\
115.79	&1036.10 &4662.73	&18	&13	&2&	10&	11&	14&	14\\
104.00	&1037.00 &4931.99	&11	&14	&9&	3&	12&	7&	12\\\hline
\end{tabular}}
\end{adjustwidth}
\end{table*}

We can perform the above computations not only for the third original
record, but for the entire original data set. We show the results
in Table~\ref{tau4}, which gives:
\begin{itemize}
\item For each original record ${\bf x}_i$, the closest 
anonymized record ${\bf y}_p$
and the permutation distance;
\item The data set-level permutation distance $d$
(this is the minimum of the record-level permutation
distances);
\item For each original record ${\bf x}_i$ 
and each anonymized attribute $Y^j$, the
variance of $S^j_i(d)$ and (between parentheses) the variance of $S^j_i(d_i)$,
where $d$ is the data set-level permutation distance ($d=1$) and 
$d_i$ is the record-level permutation distance;
\item For each anonymized attribute, the data set-level minimum variance of
the attribute values whose rank deviates
no more than $d$ from the corresponding attribute value of the 
anonymized record closest to the original record (this is the
minimum of the variances of $S^j_i(d)$).
\end{itemize}
From the data-set level permutation distance and data-set level
attribute variances, it can be seen that
{\em  the anonymized 
data set satisfies $(d,{\bf v})$-permuted privacy 
with $d=1$ and ${\bf v}=(0.01,11.07,30.26)$.}

\begin{table}
\caption{Running example. 
Minimum permutation distances, for each original record ($d_i$
for record $i$) and at the data set level ($d$).
Closest anonymized record $(Y^1,Y^2,Y^3)$  to each original record 
(record no. is given in column ${\bf y}_p$).
Anonymized attribute variances within the data set-level minimum permutation
distance $d$, for each original record (record rows) 
and at the data set level (bottom row).
Between parentheses, anonymized attribute variances
for each original record within the record-level
permutation distance $d_i$ (rather than $d$).
The anonymized data set turns out to satisfy $(d,{\bf v})$-permuted 
privacy with $d=1$ and ${\bf v}=(0.01,11.07,30.26)$.}
\label{tau4}
\begin{adjustwidth}{-3cm}{}
{\footnotesize
\begin{tabular}{|c|c|c|c|c|c|c|c|c||c|c|c|}\hline
Rec \# & $X^1$&$X^2$&$X^3$ & $Y^1$ & $Y^2$ & $Y^3$ & ${\bf y}_p$ & $d_i$ & $Var^1_i$ & $Var^2_i$ & $Var^3_i$\\\hline	
1&103.69	& 981.80& 4928.80	&108.18	&972.62	&4876.73 & 1 & 4 & 0.48 (12.45) &  69.14 (682.15)  &  388.07 (2170.53)\\
2&93.13	&980.97	&4931.16	&96.60	&1020.73 &5005.04 & 2 & 4 &6.57 (31.12) & 69.14 (682.15)   & 388.07 (2170.53)\\
3&100.87	&902.21	&5108.54	&105.26	&882.92	&4900.68 & 10 & 4 &1.63 (24.70)&    155.00 (896.76)  &  1692.52 (20167.78)\\
4&95.24	&953.37	&5084.18	&88.02	&944.54	&4949.78 & 6 & 4 &12.83 (32.47)&     64.36 (1332.66)  &  1692.52 (20167.78)\\
5&96.18	&1086.34 &5212.25	&91.57	&1057.83 &5267.57 & 5 & 2 &12.83 (16.94)&    112.36 (118.46)  &  1738.89 (14751.48)\\
6&93.16	&986.70	&5232.96	&100.41	&991.34	&5230.64 & 6 & 3 &6.57 (22.88) &    69.14 (414.40)   & 1738.99(16208.14)\\
7&95.50	&952.13	&4824.95	&100.31	&959.89	&4824.03 & 7 & 1 &12.83 (12.83) &   385.03 (385.03)  &  2612.38 (2612.38)\\
8&115.53	&988.44	&5437.43	&123.37	&1061.23&5450.70 & 17 & 4 &1.23 (18.76) &    69.14 (682.15)  &  8384.15 (2257.38)\\
9&98.99	&941.48	&4835.05	&103.12	&903.25	&4752.03 & 7 & 1  &3.14 (3.14) &   385.03 (385.03)  &  2612.38 (3407.82)\\
10&109.96	&984.87	&4950.48	&104.82	&912.77	&4997.61 & 13 & 4 &8.12 (21.96) &    69.14 (682.15)   &  555.30 (2952.80)\\
11&99.72	&1005.19 & 5158.64	&87.83	&1025.01 &5166.63 & 6 & 1 &3.14 (3.14)&   147.25 (147.25)  &  3407.82 (1676.76)\\
12&116.75	&1057.63 &4986.25	&112.21	&1082.43 &4988.44 & 12 & 4 &11.06 (13.03)&     14.43 (169.01)&    429.60 (1313.97)\\
13&107.62	&1025.13 &4954.28	&114.29	&988.93	&4889.75 & 20 & 3 &8.12 (16.70) &    41.95 (359.80)  &  555.30 (2257.38)\\
14&87.62	&1031.74 & 4905.71	&90.83	&1049.58 &4902.04& 14 & 3 &0.01 (1.47)&    29.73 (248.09)  &  208.80 (15949.39\\
15&109.81	&971.09	&4941.81	&113.64	&1002.19 &5020.71& 10 & 4 &8.12  (21.96)&   115.82 (937.01)  &  555.30 (95.84)\\
16&110.63	&1052.34 & 4495.19	&103.07	&1052.03 &4519.2 & 19 & 4 &5.34  (22.74)&    11.07 (190.00)  &  5145.91 (18406.42)\\
17&113.76	&972.20	& 4893.50	&117.00	&962.84	&5087.90 & 13 & 1 &0.75  (0.75)&   115.82 (115.82)  &  95.84 (95.84)\\
18&105.59	&1027.64 & 5143.05	&89.43	&1049.97 &5072.79 & 15& 3 & 2.22 (14.82)&     41.95 (359.80) &   3407.82 (18406.42)\\
19&108.21	&990.58	& 4714.76	&115.79	&1036.10 &4662.73 & 1 & 2 &8.12 (12.76) &    33.26 (252.94)  &  4352.91 (15949.39)\\
20&104.74	&1023.96 &4900.79	&104.00	&1037.00 &4931.99 & 20 & 2 &0.27 (2.94)&     41.95 (160.52)  &  30.26 (334.85)\\\hline
\multicolumn{8}{|c|}{Data set-level permutation distance $d$ and 
variances $Var^1, Var^2, Var^3$ }&1&0.01 & 11.07& 30.26\\\hline 
\end{tabular}}
\end{adjustwidth}
\end{table}

Obviously, the data protector, who has access to the entire original 
data set and the entire anonymized data set, can verify as
described in this section whether
the anonymized data set satisfies $(d,{\bf v})$-permuted privacy for any $d$
and ${\bf v}$ of his choice. 
The most interesting feature, however, is 
that {\em each subject can check whether $(d,{\bf v})$-permuted
privacy with respect to her original record is satisfied by the anonymized
data set for some $d$ and ${\bf v}$ of her choice}. The subject only needs
to know her original record and the anonymized data set: for example, the 
subject having contributed the third 
original record can compute the permutation 
distance as described in Table~\ref{tau3} and also compute variances
of any subset of anonymized attribute values.

\section{Intruder model in anonymization}
\label{intruder}

There are some fundamental differences between data encryption 
and data anonymization: whereas the receiver of the encrypted data 
has the key to decrypt the ciphertext back to plaintext, 
the user in anonymization has access {\em only} to what plays the role of 
the ciphertext, that is, the anonymized data.  
Consequently, while it makes sense to release encrypted data that 
disclose absolutely nothing about the underlying plaintext 
(perfect secrecy, \cite{Shannon49}), it does not make sense to 
release anonymized data that disclose absolutely nothing about the 
underlying original data. The objective of microdata release is to 
provide information to the public, which means that some disclosure 
is inherently inevitable. Even if data are anonymized prior to release, 
disclosure is still inevitable, because zero disclosure happens 
if and only if the anonymized data are completely useless, 
which makes the data release operation completely absurd. 
In fact, the privacy-first approach to data anonymization 
runs this risk of absurdity when too stringent privacy 
parameters are selected and enforced. 

Another issue that complicates matters is that any user 
of anonymized data could,
{\em potentially}, be also an intruder. 
Hence, modeling the intruder in anonymization 
is difficult since we have to 
consider many potential levels of intruder's knowledge. 
Fortunately, and in spite of the aforementioned fundamental
differences, data encryption does offer some principles that remain
useful to tackle this characterization.

In cryptography, several different attack scenarios are 
distinguished depending on the intruder's knowledge: 
ciphertext-only (the intruder only sees the ciphertext), 
known-plaintext (the intruder has access to one or more pairs of 
plaintext-ciphertext), chosen-plaintext (the intruder can choose 
any plaintext and observe the corresponding ciphertext), 
chosen-ciphertext (the intruder can choose any ciphertext 
and observe the corresponding plaintext). 

In anonymization, we can equate the original data set to 
a plaintext and the anonymized data set to a ciphertext. Hence, a 
ciphertext-only attack would be one in which the intruder has 
access only to the anonymized data: this class of attacks can be 
dangerous, as shown by~\cite{Sweeney13} for de-identified DNA data, 
by~\cite{Narayanan08} for Netflix data and by~\cite{Barbaro06} for the AOL 
data. Even if potentially dangerous, assuming that the intruder only
knows the anonymized data can be na\"{\i}ve in some situations.
For example, if the intruder is one of the subjects in the data set, 
he will normally know his own original record.

On the other hand, the strongest attacks in cryptography, 
namely chosen-plaintext and chosen-ciphertext attacks, 
assume some interaction between the intruder and the encryption
system. Thus, they are not relevant in a non-interactive 
anonymization setting such as the one we are considering (release of 
anonymized data sets). 

Hence, the strongest attack that anonymization of data sets 
must face is the known-plaintext attack. In this attack, 
one might think of an intruder knowing particular original 
records and their {\em corresponding} anonymized versions; 
however, this is unlikely, because anonymization precisely breaks 
the links between anonymized records and corresponding original records. 
A more reasonable definition for a known-plaintext attack in anonymization 
is the following.

\begin{definition}[Known-plaintext attack in anonymization]
\label{def1} 
An attack of this class is one in which the intruder knows the entire original data set (plaintext) and the entire corresponding anonymized data set (ciphertext), his objective being to recreate the {\em correct linkage} 
between the original and the anonymized records.
\end{definition}

We observe that our definition of the intruder is stronger than any 
other prior such definition in the data set anonymization scenario. 
One of the key issues in modeling the intruder in this context 
is to define his prior knowledge, including available
background knowledge. As mentioned above, 
we assume that the intruder has maximum knowledge: he knows 
${\bf X}$ and ${\bf Y}$,
from which he can recreate ${\bf Z}$ by reverse mapping; hence, 
he only lacks the key, that is, the correct linkage between 
${\bf X}$ and ${\bf Z}$. 
In particular, assuming knowledge of ${\bf X}$ by the intruder
eliminates the need to consider 
the presence/absence of external background knowledge 
(typically external identified data sets linkable through quasi-identifiers) 
when evaluating the ability of the intruder to disclose information. 
In this respect, the intruder's background knowledge 
is as irrelevant in our intruder model as it is in 
$\varepsilon$-differential privacy.


As hinted above, knowledge of ${\bf X}$ 
allows our intruder to reverse-map ${\bf Y}$ to ${\bf Z}$, even 
if the data administrator only releases ${\bf Y}$. 
Hence, using the permutation paradigm of Section~\ref{paradigm},
we can say that the intruder is able to remove the noise addition
step in the functional equivalence of 
anonymization, so that he is only confronted with permutation.
In other words, if we consider that noise addition is governed
by one key (the random seed for the noise) and permutation by 
another key (the random seed for the permutation), reverse mapping
allows the intruder to get rid of the former key and focus on the latter.

\subsection{Record linkage computation by the intruder}
\label{computation}

The concept of record linkage has a long history in the disclosure 
limitation literature. Many different record linkage procedures 
have been suggested and two of the main procedures are distance-based 
record linkage and probabilistic record linkage 
(see~\cite{Torra03} for a discussion on them). 
Yet, one of the key aspects affecting the success of record 
linkage is knowledge of the underlying procedure 
used to anonymize the data. For example, if normally distributed 
noise is used to mask the original data, then it has been 
shown that an optimal distance-based record linkage can 
be performed~\cite{Fuller93}. But in other cases, it cannot be 
shown that any particular record linkage method performs optimally. 
This results from the simple fact that  
record linkage must be able to reverse the anonymization procedure and,
with a host of different anonymization procedures, this is a challenging
task.

From the perspective of our intruder, however, all anonymization 
procedures are reduced to permutations of the original data. 
Thus, the best option to guess which original record corresponds
to which permuted record is to use the above described 
permutation distance computation algorithm with the small
adaptation of replacing the anonymized data set ${\bf Y}$ by 
the permuted data set ${\bf Z}$ in Algorithm~\ref{alg2}. 
Note that we do not preclude the intruder from using 
some other record linkage procedure and using 
Algorithm~\ref{alg2} for purposes of confirmation. 

For the sake of illustration, Table~\ref{tau5} shows the linkages
our intruder would obtain when using Algorithm~\ref{alg2} on
the ${\bf X}$ and ${\bf Z}$ data sets of our running example.
The following remarks are in order:
\begin{itemize}
\item For some records in ${\bf X}$ (record nos. 1, 9, 11 and 19), 
multiple matches are obtained. For example, 
for the first original record, both the first 
and the seventh permuted records are at shortest permutation distance.
\item Some records in ${\bf Z}$ 
(record nos. 4, 7, 10, 12 and 13) are matches to multiple
records in ${\bf X}$, whereas some records in ${\bf Z}$ 
(record nos. 3, 8, 16 and 18) are
matches to no record in ${\bf X}$.
\end{itemize}
The intruder can realize the above, which diminishes his confidence
in the accuracy of the re-identification process. 

Furthermore,
it can be seen in Table~\ref{tau5} 
that 5 records are correctly linked, 4 records have multiple
matches and the remaining 11 records are misidentified.
While the data protector can realize this, {\em the intruder cannot
tell with certainty correct linkages from misidentifications},
because he does not know the correct linkages.
The data protector may use the proportion of correct linkages as a 
metric to evaluate the protection provided by anonymization.

\begin{table}
\caption{Running example. Record linkages computed by the intruder.
For each original record in ${\bf X}$ (record no. specified
in column $\#X$), permuted records in ${\bf Z}$ at shortest permutation
distance (record no. (or nos.) specified in column $\#Z$). 
Both for ${\bf X}$ and ${\bf Z}$, the values of record no. $i$ can
be found in the $i$-th row of Table~\ref{tau2}.}
\label{tau5}
\begin{center}
\begin{tabular}{|c|c|c|}\hline
$\#X$ & $\#Z$ & $dz_i$ \\\hline	
1 &  1, 7 & 4	\\
2 &  4 	& 3\\
3 &  10 	& 3\\
4 &  4 	& 4 \\
5 &  5	& 2\\
6 &  11	& 2\\
7 &  7	& 2 \\
8 &  17	& 5\\
9 &  7, 9	 & 3\\
10 & 15	& 3\\
11 & 2, 6 & 4	\\
12 & 12	 & 5 \\
13 & 20	& 3\\
14 & 14	& 3 \\
15 & 10	& 3\\
16 & 19	& 5\\
17 & 13	& 2\\
18 & 12 & 5	\\
19 & 13, 19 & 4	\\
20 & 20	& 3\\\hline
\end{tabular}
\end{center}
\end{table}

\subsection{Record linkage verification by the intruder}
\label{verify}

The inability of an intruder to assess the accuracy of 
re-identification via record linkage is often viewed 
as providing plausible deniability to the data protector.
In other words, even if the intruder boasts the 
record linkages he has computed (something like Table~\ref{tau5}),
he cannot prove {\em with certainty}  which linkages are correct. Hence, any
subject seeing that she has been correctly re-identified
by the intruder ({\em e.g.}, the subject behind original
record no. 4 in Table~\ref{tau5})
could be reassured by the data protector that 
re-identification has occurred by chance alone without
the intruder really being sure about it.

However, an intruder with the knowledge 
specified in Definition~\ref{def1} 
can perform the analysis described in this section 
{\em to verify how likely it is for his computed record linkages
to be correct}. 
To do this, the intruder simply needs to generate a random
set of values by drawing from the original data and then determine
the permutation distance at which a match occurs from this random data.

For instance, assume that the intruder randomly draws one value from $X^1$, 
another value from $X^2$ (independent of the draw from $X^1$), 
and a third value from $X^3$ (independent of the draws from $X^1$ and $X^2$).
Assume that the first draw yields the value of $X^1$ in the fifth original
record, the second draw the value of $X^2$ in the 19-th original record
and the third draw the value of $X^3$ in the 10-th original record.
The synthetic record formed by the intruder is 
${\bf a}=(x^1_5,x^2_{19},x^3_{10}) = (96.18, 990.58, 4950.48)$.
This record does not exist in the original data set ${\bf X}$. But even 
for this synthetic record there is some permutation distance at 
which the intruder is likely to find a matching record in {\bf Z}. 
When the permutation distance computation algorithm is used for this 
synthetic record, a match is found at distance 2 and the matched 
record in ${\bf Z}$ is the second record $(96.18, 988.44, 4986.25)$ 
(see the records of ${\bf Z}$ in Table~\ref{tau2}).

Given that the size of our running example is small, 
the intruder can perform the above analysis (computing the permutation
distance of the match in ${\bf Z}$) 
for all possible $8000$ ($=20^3$)
records resulting from three random draws. Let ${\bf A}$ 
be the data set containing these $8000$ possible records.
Within ${\bf A}$, 20 records are the original records in ${\bf X}$, 
20 the permuted records in ${\bf Z}$, and the remaining are actual synthetic 
records. Hence, for the 20 records in ${\bf Z}$, a match in ${\bf A}$ 
would be found at a permutation distance of zero. Table~\ref{tau6}
shows the distribution of the permutation distance for the 20 original records
in ${\bf X}$ and for the $8000$ possible records in ${\bf A}$. Figure~\ref{fig1}
is a graphical representation of both distributions.

\begin{table}
\caption{Running example. Distributions of the permutation distance
of the match in ${\bf Z}$ for the set ${\bf X}$ of 20 original records 
and for the set ${\bf A}$ of the $8000$ possible records that can be obtained
by respective random draws from $X^1$, $X^2$ and $X^3$.} 
\label{tau6}
\begin{center}
\begin{tabular}{|c|c|c|}\hline
Distance & Frequency & Frequency \\
$d$  & for ${\bf X}$  &   for ${\bf A}$ \\ \hline	
0 &  0.0000 & 0.0025	\\
1 &  0.0000 	& 0.0586\\
2 &  0.2000	& 0.1899\\
3 &  0.4000 	& 0.3014 \\
4 &  0.2000	& 0.2595\\
5 &  0.2000	& 0.1288\\
6 &  0.0000	& 0.0428 \\
7 &  0.0000	& 0.0143\\
8 &  0.0000	 & 0.0024\\
9 &  0.0000	& 0.0000\\
10 & 0.0000     & 0.0000\\\hline
\end{tabular}
\end{center}
\end{table}

\begin{figure}
\begin{center}
\includegraphics[width=\textwidth]{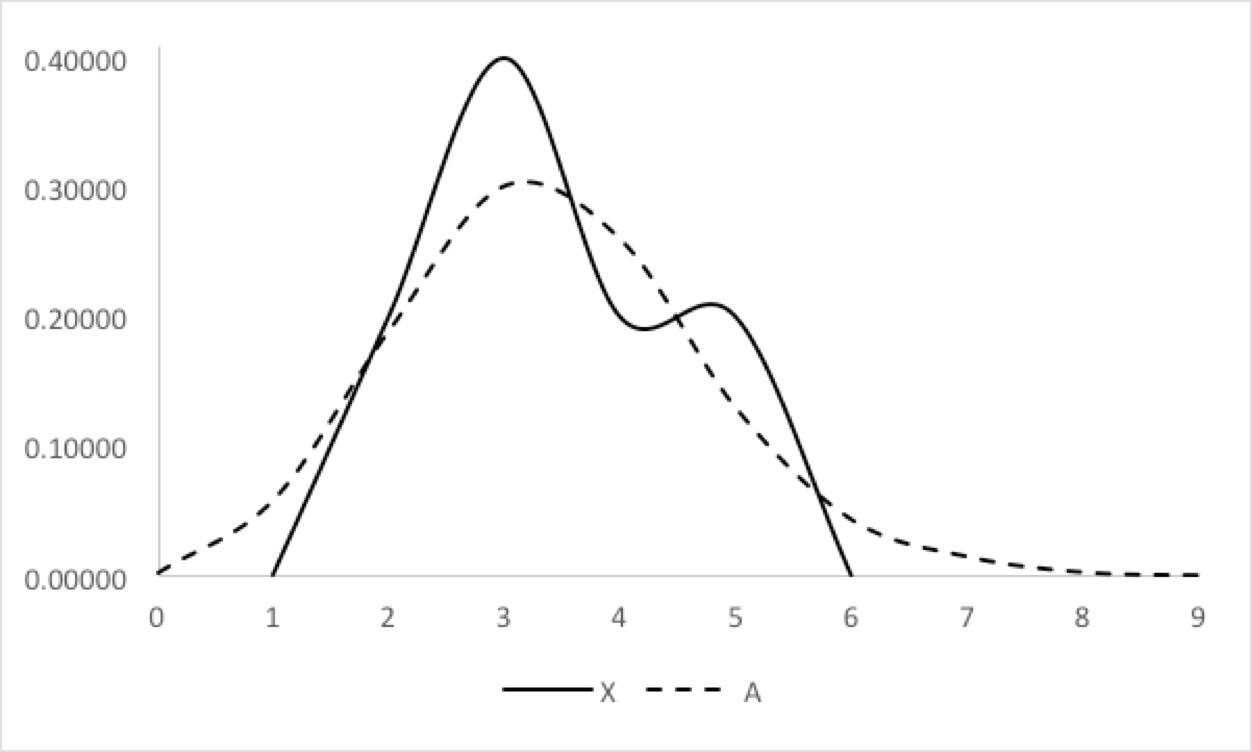}
\end{center}
\caption{Running example. Graphical
representation of the distributions of the permutation distance
of the match in ${\bf Z}$ for the set ${\bf X}$ of 20 original records
and for the set ${\bf A}$ of the $8000$ possible records that can be obtained
by respective random draws from $X^1$, $X^2$ and $X^3$.}
\label{fig1}
\end{figure}

Both Table~\ref{tau6} and Figure~\ref{fig1} highlight that the 
probability of finding a match at a particular permutation distance
for an original record in ${\bf X}$ is quite similar to the probability
of finding a match at the same permutation distance for a random
record in ${\bf A}$.
Otherwise put, when a matching record is found for an original record 
in ${\bf X}$, there is a high probability that the match
occurred by chance alone. Hence, upon seeing Table~\ref{tau6} and/or
Figure~\ref{fig1}, the intruder realizes that he 
cannot claim success in his record linkages because they
are not reliable. In conclusion, the anonymization withstands
a known-plaintext attack as per Definition~\ref{def1}. 
And, since the intruder of Definition~\ref{def1} has maximum
knowledge, the anonymized data are also safe from record linkage
by any other intruder.

As Figure~\ref{fig1} illustrates, for a very small data set
(such as the one in our running example),
even a small level of permutation is likely to 
prevent the intruder from claiming success with re-identification. 
Note that random matches occur already at distances 0, 1, 2, etc.,
so short-distance matches actually due to small anonymization 
permutation are plausible
as random matches. 
This need not be the case with larger data sets: if 
the number of records or the number of attributes are 
greater, then random matches 
at short distances may be extremely rare or even non-existing,
 in which case short-distance matches due to small permutation
are no longer plausible as random matches.
Hence, anonymizing with a 
small level of permutation may not suffice for larger data sets.

For the sake of illustration, 
consider an original data set ${\bf X}$ with $1000$ records randomly generated in 
the same way as the original data set in our running example.  
We first use perturbation through additive noise with the 
same characteristics as the one used in our running example
and we get an anonymized data set ${\bf Y}$.
Then the intruder reverse-maps ${\bf Y}$ to get 
${\bf Z}$.
While it would still be feasible to generate all potential combinations of values 
from ${\bf X}$ ($1000^3$), for purposes of computational efficiency, 
we assume the intruder generates a data set ${\bf A}$ with 
$10,000$ synthetic records by randomly and independently selecting values 
from attributes $X^1$, $X^2$ and $X^3$. 
Figure~\ref{fig2} depicts the distributions of the permutation distance
for the original records (in ${\bf X}$) and for the random records
(in ${\bf A}$). It turns out that both distributions
are practically indistinguishable. So, we are in a similar
situation as in Figure~\ref{fig1}, although
the permutation distances are much larger in 
Figure~\ref{fig2}. 
Hence, if the intruder were to 
find a match, there is a high probability that the match could
have occurred at random. We can conclude that the anonymization procedure
used to obtain ${\bf Y}$
withstands a known-plaintext attack.

\begin{figure}
\begin{center}
\includegraphics[width=\textwidth]{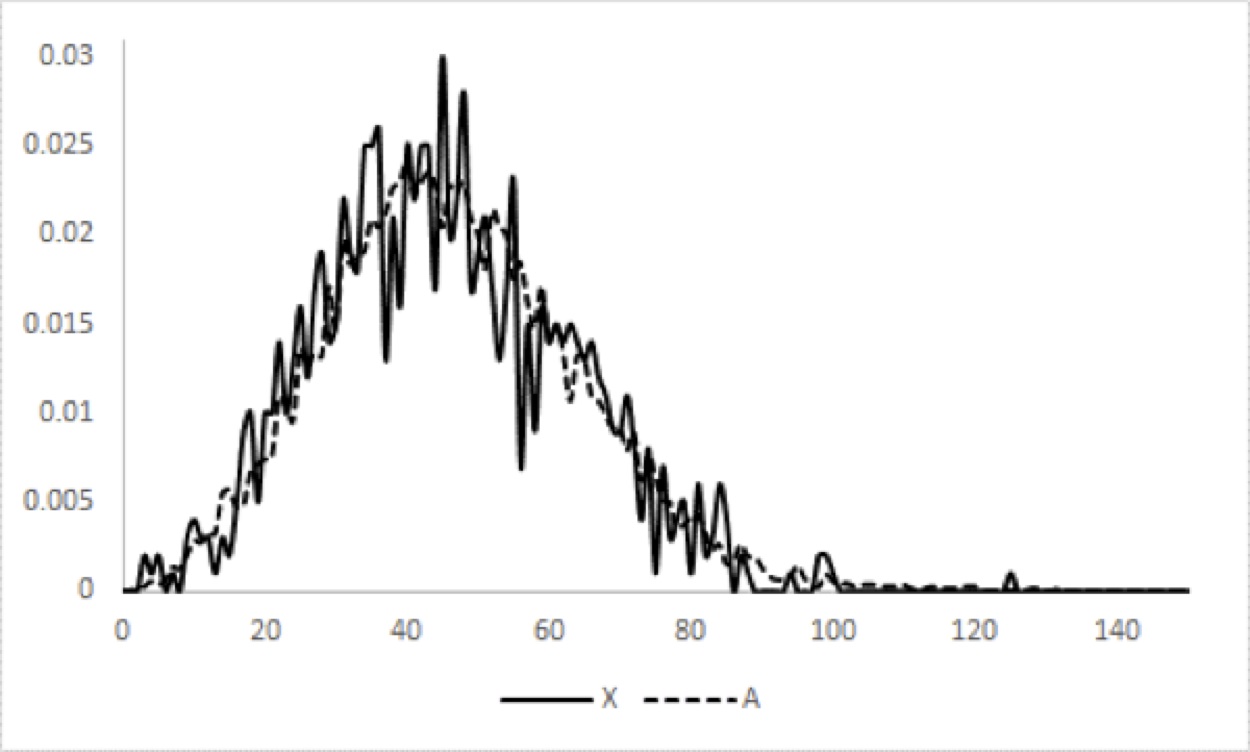}
\end{center}
\caption{Original data set ${\bf X}$ with $1000$ records
anonymized by adding $N(0,5^2)$ noise to $X^1$, $N(0,25^2)$ noise
to $X^2$ and $N(0,100^2)$ to $X^3$. Graphical
representation of the distributions of the permutation distance
of the match in ${\bf Z}$ for ${\bf X}$
and for a set ${\bf A}$ of $10,000$ random records obtained
respective and independent random draws from $X^1$, $X^2$ and $X^3$.}
\label{fig2}
\end{figure}

In contrast, consider now the same data set ${\bf X}$ with 
$1000$ records, but assume that the noise added to get
the anonymized data ${\bf Y}$ is very small.
Specifically, the noise $E^1$ is sampled from a $N(0,0.05^2)$, the 
noise $E^2$ added to $X^2$ from a $N(0,0.25^2)$ and the noise 
$E^3$ added to $X^3$ from a $N(0,1)$. Figure~\ref{fig3} and 
Table~\ref{tau7} show
the distributions 
of the permutation distance
for the original records (in ${\bf X}$) and for the random records
(in ${\bf A}$).  

\begin{figure}
\begin{center}
\includegraphics[width=\textwidth]{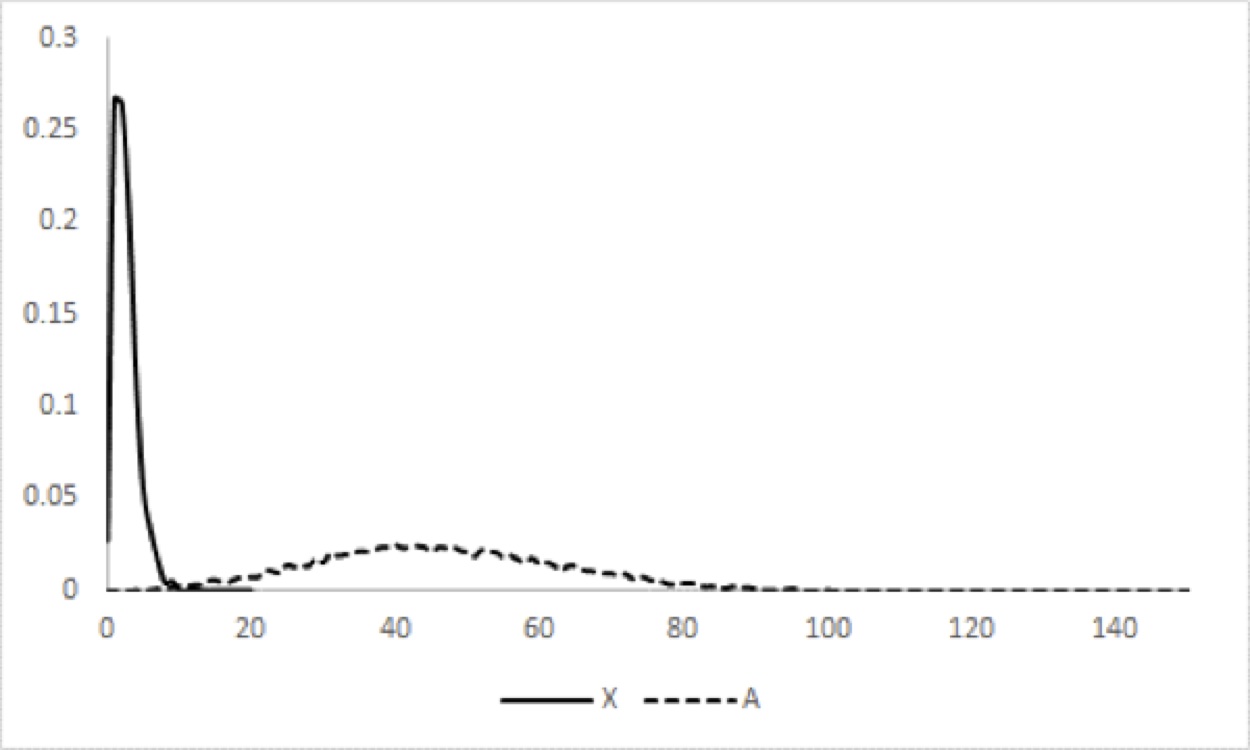}
\end{center}
\caption{
Original data set ${\bf X}$ with $1000$ records
anonymized by adding $N(0,0.05^2)$ noise to $X^1$, $N(0,0.25^2)$ noise
to $X^2$ and $N(0,1)$ to $X^3$. Graphical
representation of the distributions of the permutation distance
of the match in ${\bf Z}$ for ${\bf X}$
and for a set ${\bf A}$ of $10,000$ random records obtained
respective and independent random draws from $X^1$, $X^2$ and $X^3$.
}
\label{fig3}
\end{figure}

\begin{table}
\caption{
Original data set ${\bf X}$ with $1000$ records
anonymized by adding $N(0,0.05^2)$ noise to $X^1$, $N(0,0.25^2)$ noise
to $X^2$ and $N(0,1)$ to $X^3$. Distributions of the permutation distance
of the match in ${\bf Z}$ for ${\bf X}$
and for a set ${\bf A}$ of $10,000$ random records obtained
respective and independent random draws from $X^1$, $X^2$ and $X^3$.
}
\label{tau7}
\begin{center}
\begin{tabular}{|c|c|c|}\hline
Distance & Frequency & Frequency \\
$d$  & for ${\bf X}$  &   for ${\bf A}$ \\ \hline
0 &  0.0270 & 0.0000    \\
1 &  0.2670     & 0.0000\\
2 &  0.2650     & 0.0001\\
3 &  0.2090     & 0.0002\\
4 &  0.1170     & 0.0005\\
5 &  0.0550     & 0.0003\\
6 &  0.0330     & 0.0006 \\
7 &  0.0170     & 0.0013\\
8 &  0.0040      & 0.0012\\
9 &  0.0050     & 0.0019\\
10 & 0.0000     & 0.0028\\
11 & 0.0000     & 0.0027\\
12 & 0.0010     & 0.0032\\
13 & 0.0000     & 0.0032\\
14 & 0.0000     & 0.0054\\
$\geq 15$ & 0.0000 & 0.9766\\ \hline
\end{tabular}
\end{center}
\end{table}

From Table~\ref{tau7}, a match that occurs at a permutation distance of 0 must be
a correct match: that is, the noise added to anonymize
was so small that it did not result in
re-ordering any of the attributes. Table~\ref{tau7} also
shows that there is only a probability 0.0011
(roughly 1 over 1000) that a match is random given that its
permutation distance is $\leq 5$. In fact, 
{\em comparing the distributions 
of the permutation distance of matches for 
the original data and the random data is 
an excellent tool for the intruder to verify
on a record by record basis
how accurate his record linkages are}. 
Given the very little overlap of the distributions
shown in Figure~\ref{fig3}, the intruder can conclude
that his matches are very likely to be correct ones. In this case,
the anonymization procedure fails to withstand a known-plaintext attack.

{\em The above assessment by the intruder can also be made
by the data protector before releasing the data, in order 
to determine the optimum amount of permutation that anonymization
should introduce.}

The distribution of the permutation distance for the original values 
in ${\bf X}$ is a direct function only of the level of anonymization 
---the higher the modification introduced by anonymization,
the longer the permutation distance. The distribution of 
the permutation distance for random records grows with 
the number of records, grows with the number of attributes
and, by construction, it is independent of the anonymization method and level 
of anonymization used (the random data set contains all possible
permutations of the original records or a random large subset of them). 
A comprehensive discussion of the exact
characteristics of the distribution of the permutation distance
is beyond the scope of this paper.

To the best of our knowledge, ours is the first attempt to 
present a principled algorithm for the intruder to 
evaluate the effectiveness of the re-identification process.
Prior assessment of re-identification could only be carried out 
by the data protector and it only focused on 
the percentage of misidentifications and the percentage 
of multiple matches (in line with the analysis made in 
the last paragraph of Section~\ref{computation} above).
Further developments may allow the intruder 
to assess the extent to which the two distributions are different 
(by using measures such as Hellinger's distance) or develop 
formal statistical tools by treating the distribution of the match distance 
for the random records as the distribution of the statistic under the 
null hypothesis in hyphotesis testing. 

Finally, note that {\em when 
the anonymization method involves only permutation without noise addition} 
(which is the case with data swapping~\cite{Dalenius82} 
and data shuffling~\cite{Muralidhar06}),
{\em a data subject with access to just her own record in ${\bf X}$ 
can not only
learn the permutation distance $d$ of her record} (as described
in Section~\ref{permuted}), {\em but she can also
verify whether $d$ is safe}. To this end,
the subject generates ${\bf A}$ from the masked data ${\bf Z}$
(${\bf Z}$ can be used instead of ${\bf X}$, because one 
data set is a permutation of the other) and then checks
whether a match at distance $d$
is plausible as a random match; if yes, then $d$
is safe. One may assume that the data protector
has checked that the permutation distance of all records is safe, but
giving each subject the possibility to check it is an attractive
feature of pure-permutation anonymization.

\section{Anonymization transparency towards the user}
\label{transparency}

In this section, we discuss the user in the context of the permutation-paradigm
of anonymization presented in Section~\ref{paradigm}.
There is one tenet from data encryption that can be usefully applied 
to data anonymization: Kerckhoff's principle, which states that 
the encryption algorithm must be entirely public, with the key being 
the only secret parameter. Nowadays,
statistical agencies and other data releasers often refrain from publishing 
the parameters used in anonymization (variance of the added noise, 
proximity of swapped values, group size in microaggregation, etc.). The 
exception is when the privacy-first approach is used (based on a privacy model), 
in which case the anonymization parameters are explicit and dictated by 
the model. However, as mentioned above, most real data releases are 
anonymized under the utility-first approach. Withholding the parameters 
of anonymization is problematic for at least two reasons:
\begin{enumerate}
\item The legitimate user cannot properly evaluate the utility of the 
anonymized data.
\item Basing disclosure protection on the secrecy of the anonymization 
algorithm and its parameterization is a poor idea, as it is hard to 
keep that much information secret and it is better to expose algorithms
and parameterizations to public scrutiny to detect any weaknesses in them.
\end{enumerate}

One might argue that the parameters of an anonymization method play 
the role of the key in cryptography and must therefore be withheld. 
We contend that this is a wrong notion, because whereas cryptographic 
keys are randomly chosen, anonymization parameters are not 
(there are typical values for noise variance, etc.). The most 
similar thing to a cryptographic key in the context of anonymization 
are the random seeds used by (pseudo-)randomized anonymization methods.  

It is also important to note that Kerckhoff's principle is of no 
consequence to the intruder modeled according to cryptographic principles. 
According to our definition of the intruder, the 
anonymization method and the level of anonymization play no role 
in the re-identification process. Once the intruder has performed 
reverse mapping, the only remaining unknown is the random key 
used for permuting the values. And {\em we have shown how the intruder 
can best guess the permutation used} (Section~\ref{computation}) 
{\em and then evaluate the accuracy of his guess without 
any information about the anonymization mechanism} 
(Section~\ref{verify}). 
Hence, for our intruder, the claim that following Kerckhoff's 
principle will result in increased disclosure risk is incorrect. 
Actually, following Kerckhoff's principle harms
none of the stakeholders in the microdata release 
(data protector, subject, intruder and user)
and it is extremely valuable for the user.
For these reasons, we believe that the data protector 
must always release details about the anonymization
methods and parameters used. We formalize this notion
in Definition~\ref{def2}. 

\begin{definition}[Anonymization transparency to the data user] 
\label{def2}
An anonymization method is said to be transparent to the data user
when the user is given all details of the anonymization except the 
random seed(s) (if any are used for pseudo-randomization).
\end{definition}

\section{Conclusions and future research}
\label{conclude}

We have presented a new vision of microdata anonymization 
that opens several new directions.
 
First, we have shown how knowledge of the values of the
original attribute allows reverse mapping the values of the 
anonymized attribute into a permutation of the original attribute
values. This holds for any anonymization method and for any
attribute whose values
can be ranked (and in fact any data are amenable to some sort of ranking).
Hence, any anonymization method can be viewed as a permutation
followed by a (small) noise addition.
This vision applies to any anonymization
method and it allows easily comparing methods in 
terms of the data utility and the privacy they provide.

Based on the permutation plus noise paradigm, 
we have stated a new privacy 
model, called $(d,{\bf v})$-permuted privacy, that focuses
on the minimum permutation distance achieved and on the variance of the attribute
values within that distance.
The advantage of this privacy model with respect to previous 
methods in the literature is that it is not only verifiable
by the data protector, but also by each data subject having
contributed a record to the original data set (subject-verifiability).

Then we have precisely defined a maximum-knowledge adversarial model
in anonymization. Specifically, we have shown how our intruder
can best guess the permutation achieved by an anonymization method 
and how he can assess the quality of his guess. The intruder's 
assessment is independent of the anonymization method used 
and it also tells the data protector the right
level of permutation needed to protect against re-identification.

Regarding the data user, we have argued why 
Kerckhoff's assumption should be the 
rule in anonymization, just as it is the rule in encryption.
Releasing the details of anonymization introduces no weakness 
and it is extremely useful to the user.
This calls for anonymization that is transparent to the user.

We have illustrated the concepts 
and procedures introduced throughout the paper with a running example.

This paper opens
a great number of future research lines. These include the following:
\begin{itemize}
\item Extend the reverse-mapping conversion of Algorithm~\ref{alg1}
for any type of attribute (that is, nominal in addition to numerical 
or ordinal). 
\item Explore the consequences of the permutation
paradigm of anonymization for local perturbation methods.
\item Regarding the adversarial model, rigorously
characterize the distribution of the permutation distance
and tackle the issues sketched at the end of Section~\ref{verify}.
\item In line with the cryptography-inspired model of anonymization, seek 
information-theoretic measures of anonymity focused on 
the mapping between the original and anonymized records output by a specific anonymization method. 
\item Produce an inventory of anonymization methods in the literature
that are transparent to the data user according to Definition~\ref{def2}.
In particular, investigate to what extent 
deterministic methods (using no randomness seeds, {\em e.g.}, microaggregation, coarsening, etc.) can be transparent.
\end{itemize}

\section*{Disclaimer and acknowledgments}

The following funding sources are gratefully acknowledged:
Government of Catalonia (ICREA Acad\`emia Prize to the
first author and grant 2014 SGR 537),
Spanish Government (project TIN2011-27076-C03-01 ``CO-PRIVACY''),
European Commission (projects
FP7 ``DwB'', FP7 ``Inter-Trust'' and H2020 ``CLARUS''),
Templeton World Charity
Foundation (grant TWCF0095/AB60 ``CO-UTILITY'')
and Google (Faculty Research Award to the first author).
The first author is with the UNESCO Chair in Data Privacy.
The views in this paper are the authors' own and
do not necessarily reflect
the views of UNESCO, the Templeton World Charity
Foundation or Google.

\end{document}